\documentclass[aps, prb, superscriptaddress, showpacs, twocolumn, longbibliography]{revtex4-1}
\makeindex
\usepackage{amsmath}
\usepackage{amsthm}
\usepackage{amssymb}
\usepackage{mathrsfs}
\usepackage{bm}
\usepackage{float}
\usepackage[pdftex]{graphicx}
\usepackage[utf8]{inputenc}
\usepackage{xcolor}
\usepackage{mathtools}
\usepackage{comment}
\usepackage{ulem} 
\usepackage{xcolor}

\providecommand{\includegraphics}[2][width=\textwidth]{$#2$}

\definecolor{citecol}{rgb}{0.0, 0.6, 0.0}
\definecolor{linkurl}{rgb}{0.2, 0.2, 0.8}
\usepackage[colorlinks=true, urlcolor=linkurl,citecolor=citecol]{hyperref}

\newcommand{\DM}{Dzyaloshinskii-Moriya }
\def\H{\mathcal H}

\def\Tr{{\rm Tr}}

\begin{document}

\title{Effect of perturbations on the kagome $S=1/2$ antiferromagnet at all temperatures}
\author{Bernard Bernu}
\affiliation{Sorbonne Universit\'e, CNRS, Laboratoire de Physique Th\'eorique de la Mati\`ere Condens\'ee, LPTMC, F-75005 Paris, France}
\email{bernu@lptmc.jussieu.fr}
\author{Laurent Pierre}
\affiliation{Paris X, Nanterre}
\author{Karim Essafi}
\affiliation{Sorbonne Universit\'e, CNRS, Laboratoire de Physique Th\'eorique de la Mati\`ere Condens\'ee, LPTMC, F-75005 Paris, France}
\author{Laura Messio}
\affiliation{Sorbonne Universit\'e, CNRS, Laboratoire de Physique Th\'eorique de la Mati\`ere Condens\'ee, LPTMC, F-75005 Paris, France}
\affiliation{Institut Universitaire de France (IUF), F-75005 Paris, France}
\email{messio@lptmc.jussieu.fr}

\date{\today}
\pacs{
	02.60.Ed	
	05.70.-a	
	71.70.Gm	
	75.10.jm	
	75.40.Cx	
	02.70.Rr	
}

\begin{abstract}
	The ground state of the $S=1/2$ kagome Heisenberg antiferromagnet is now recognized as a spin liquid, but its precise nature remains unsettled, even if more and more clues point towards a gapless spin liquid. 
	We use high temperature series expansions (HTSE) to extrapolate the specific heat $c_V(T)$ and the magnetic susceptibility $\chi(T)$ over the full temperature range, using an improved entropy method with a self-determination of the ground state energy per site $e_0$. 
	Optimized algorithms give the HTSE coefficients up to unprecedented orders (20 in $1/T$) and as exact functions of the magnetic field. 
	Three extrapolations are presented for different low-$T$ behaviors of $c_V$: exponential (for a gapped system), linear or quadratic (for two different types of gapless spin liquids). 
	We study the effects of various perturbations to the Heisenberg Hamiltonian: Ising anisotropy, \DM interactions, second and third neighbor interactions, and randomly distributed magnetic vacancies.
	We propose an experimental determination of $\chi(T=0)$, which could be non zero, from $c_V$ measurements under different magnetic fields. 
\end{abstract}

\maketitle

\section{introduction}
The physics of the spin $S=1/2$ kagome lattice, with first neighbor Heisenberg antiferromagnetic interactions\cite{MENDELS2016455} (KHAF) has recently known two major progresses. One is experimental, with the realization of high quality crystals of Herbersmithite\cite{PhysRevB.83.100402}, opening the possibility of precise measurements\cite{Nature_DM, Mendels}; the other is numerical, with the understanding of the bias tending to erroneously favor a gapped spin liquid (SL) ground state in DMRG simulations\cite{He2017,Liao2017, PhysRevB.97.104401}. 
A gapless SL ground state is now almost a consensus, supported by recent precise measurements of the low-$T$ magnetic susceptibility\cite{Mendels}. 
However, there remain several types of bidimensional gapless SL, among which the $U(1)$ SL (with a ponctual Fermi surface) and the Fermi SL (with a linear Fermi surface)\cite{PhysRevLett.98.117205}. 
They distinguish themselves notably by the low-$T$ behavior of their specific heat: 
a linear behavior, $c_V \propto T$, is a characteristic of a Fermi SL, whereas a quadratic one, $c_V\propto T^2$, is an indication of a $U(1)$ SL (to compare to $c_V \propto T^2 e^{-\Delta/T}$ for a gapped phase, where $\Delta$ is the gap). 
We label these different cases by an integer $\alpha=1$ or 2 in the gapless cases ($c_V\propto T^\alpha$) and $\alpha=0$ in the gapped case. 
Up to now, neither the experimental nor the theoretical works are able to determine $\alpha$ for the KHAF, even if recent theoretical and experimental results seem to indicate that $\alpha\ne0$. 

But all these considerations pre-suppose that Herbertsmithite is effectively described by a KHAF on perfect and independent kagome planes. 
In reality, this model suffers from several perturbations: dilution, Ising anisotropy, \DM (DM) interactions, further neighbor interactions...
Previous studies show that their effects on the ideal Hamiltonian are moderate: whatever the phase of the KHAF ground state, it seems stable for small values of these perturbations. 
However, they can quantitatively influence the finite temperature thermodynamic measurements. 
Thus, we use in this article high temperature series expansions (HTSE) to explore the finite temperature effects of a magnetic field $h$ and of all the previously listed perturbations in the three cases $\alpha=0$, 1 or 2. 
It illustrates the difficulty to fit experimental data for many free parameters and without knowing $\alpha$. 
However, we extract from all these results a way to determine the zero temperature magnetic susceptibility $\chi_0$ from $c_V$ measurements under different $h$, and we anticipate the synthesis of parent compounds of Herbertsmithite with tunable perturbations to KHAF. 

HTSE exactly calculates the Taylor coefficients of thermodynamic quantities in powers of the inverse temperature $\beta=1/T$. 
From these coefficients, one can reliably and easily reconstruct the quantities from infinite down to moderate temperatures of the order of the interaction strength, using either the raw series, Pad\'e approximants (PAs), or methods as differential Pad\'e approximants, Euler transformation, \dots\cite{Oitmaa1996, Roger, Lohmann2014, Hehn2017}.
When there is no singularity down to $T=0$ in the thermodynamic functions (i.e. no phase transition, as notably in SL phases), it is possible to extrapolate HTSE over the full range of temperature. 
In this case, the entropy method combines HTSE with an hypothesis on $\alpha$ to get thermodynamic quantities as the specific heat per site $c_V$ or the magnetic susceptibility per site $\chi$\cite{Bernu2001, Bernu2015, PhysRevE.95.042110}.
This method, thereafter denoted HTSE$+s(e)$, is fully relevant to extract the Hamiltonian parameters from experimental results\cite{PhysRevB.68.113409, PhysRevLett.118.237203, Kapellasite_HT, Kapellasite_cuboc2}.  

We get in this article HTSE coefficients up to an order notably larger than previously\cite{Bernu2015}, in the presence of all the above interactions and with an exact dependency in $h$. 
Moreover, we present here the extrapolations on the KHAF supposing a \textit{gapless} spin liquid, with a special emphasis on $\chi$ and $c_V$ (see also ref.\cite{Misguich2005} on $c_V$), which can be measured experimentally\cite{Mendels}. 
These extrapolations require input parameters: $\alpha$, the ground state energy per site $e_0$ and $\chi_0$.
Often, $\alpha$ is known, as for Néel or gapped ground state.
Except for ferromagnetic states, $e_0(h)$ and $\chi_0=-d^2e_0/dh^2$ are usually unknown.
We present here a new method of self-determination of $e_0$ that overcome this obstacle. 
In Sec.~\ref{sec:rawHTSE}, we present the results of raw series. 
We  then discuss the extrapolation method and present the results on the perfect KHAF in Sec.~\ref{sec:extrapolation}. 
Sec.~\ref{sec:perturbations} is devoted to the study of several perturbations, followed by the effects of a magnetic field $h$.
Concluding remarks are in the last section. 
Supplemental material\cite{suppMat} gives details on HTSE$+s(e)$ and furnish more figures illustrating the effect of the perturbations. 

\section{Raw HTSE coefficients with exact dependency in \texorpdfstring{$h$}{h}}
\label{sec:rawHTSE}
We first focus on the raw series of the thermodynamic limit of the logarithm of the partition function $\lim_{N\to\infty}\frac{\ln Z}N$ in powers of $\beta$  with, as first main result of this article, their obtention as exact functions of $h$. 

The KHAF Hamiltonian $\H$ consists in spins $S=1/2$ on a kagome lattice, in presence of an arbitrary magnetic field $h$ (times a factor $g\mu_B$, set to 1 in the following), with antiferromagnetic interactions on all pairs of nearest neighbors:
\begin{equation}\label{EQ-H0}
\H_0 = J_1 \sum_{\langle i,j\rangle} {\bm S}_i\cdot {\bm S}_j,\qquad 
  \H = \H_0 - h S^z,
\end{equation}
where $S^z = \sum_i S^z_i$ is the total spin along the $z$ direction and $\mathbf S_i$ the spin operator on site $i$. 
$J_1$ is set to unity in the following. 
The partition function is:
\begin{align}
  Z=\Tr\,e^{-\beta \H} = \sum_{n=0}^\infty \frac{(-\beta)^n}{n!}\Tr(\H^n). 
 \end{align}
After keeping the part of the traces $\Tr(\H^n)$ originating from connected clusters with $n$ links on the lattice,
it gives us the following HTSE in powers of $\beta$, where coefficients are finite order polynomials of $h^2$: 
\begin{align}
\label{eq:lnZ}
\lim_{N\to\infty}\frac {\ln Z}N&=\ln 2+\sum_{n=1}^\infty \left(\sum_{k=0}^{n/2} Q_{n,k} h^{2k}\right)\beta^n.
\end{align}
The first coefficients $Q_{n,0}$ and $Q_{n,1}$ are related to the HTSE of respectively $c_V$ and $\chi$ at $h=0$, and are the only ones that were calculated up to now\cite{Bernu2001, Misguich2005, Bernu2015}: the effects of a finite $h$ were unaccessible (some further terms were calculated for other models\cite{doi:10.1143/JPSJ.35.25, Bernu2001}, without being exploited or still strongly limiting the possible values of $h$). 

Beside the now exact treatment of $h$, we get access to unprecedented orders despite the exponential complexity of the calculations. 
$Q_{n,k}$ are determined for $n$ up to 20, against 17 previously \cite{Bernu2015}.
Fig. \ref{fig:Cv_chi} shows that the raw HTSE diverges below T=1, while the PAs converge down to 0.5 allowing to describe the main peak of $c_V$.

\section{Extrapolation over the full temperature range}
\label{sec:extrapolation}
In the thermodynamic limit, canonical and micro-canonical ensembles are equivalent. 
It implies that the information contained in $Z(T,h)$ is the same as in the entropy per spin $s(e,h)$, with $e$ the energy per spin. 
At fixed $h$, $s$ and $e$ are monotonous functions of $T$, going from 
$e_0(h)$ and $s_0=0$ at $T=0$,
to
$e_\infty=0$ and $s_\infty=\ln(2S+1)$ at $T=\infty$. 
These constraints near $e_0(h)$ are equivalent to the two sum rules on $c_V$, but more easily imposed on $s(e,h)$\cite{Bernu2001}. 
Moreover, the behavior of $s(e,h)$ for $e\to e_0(h)$ can be infered from the (known or supposed) low energy properties of the model. 
Thus, we work in the micro-canonical ensemble\cite{Bernu2001,Misguich2005,Bernu2015}.
From the HTSE, Eq.~\eqref{eq:lnZ}, we deduce the series expansion of $s(e, h)$ around $e_\infty$ and extrapolate this function over the full interval $\lbrack e_0(h),e_\infty\rbrack$. 
To remove the singularity of $s$ at $e_0$, we introduce an auxiliary function $G_\alpha(s(e, h))$. 
Then, PAs of this function of $e$ are used to reconstruct $s$\cite{suppMat}. 

\begin{figure}[t]
	\includegraphics[width=.48\textwidth]{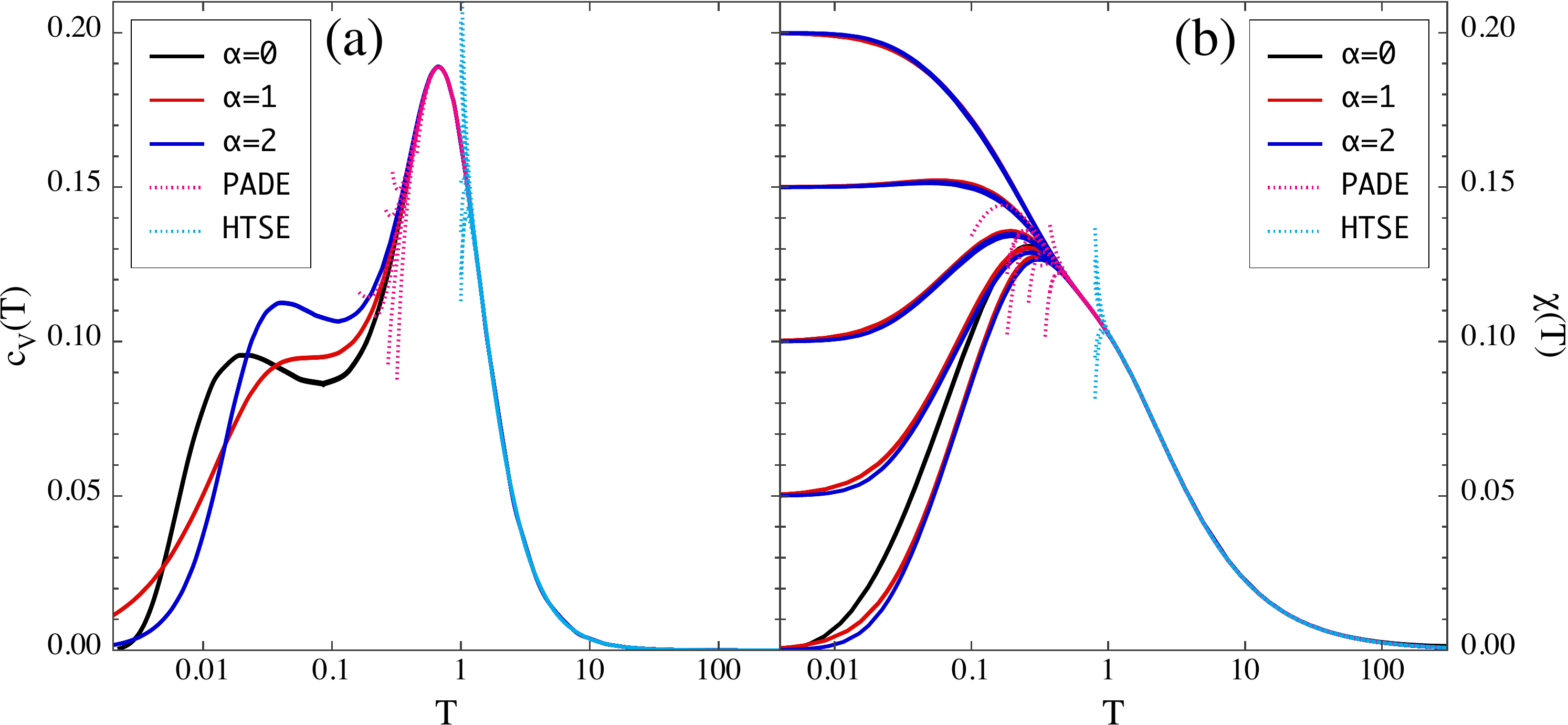}
	\caption{Results from HTSE$+s(e)$ at order 20 in $\beta$ for $h=0$ on the KHAF. 
		$(a)$ Specific heat $c_V$; 
		Gapped ($\alpha=0$, black lines) and gapless ground states ($\alpha = 1$, red lines, and $2$, blue lines) are considered and $e_0$ is fixed to -0.4372, -0.4384 and -0.4395 respectively.
		Dashed cyan lines are the raw HTSE of orders 13 through 20.
		Dashed magenta lines are the PAs $d=6$ through 14 of HTSE at order 20.
		$(b)$ Same as in $(a)$ for the magnetic susceptibility $\chi$. 
        Several scenario for the $\chi_0$ value are presented.
		\label{fig:Cv_chi}
	}
\end{figure}
This HTSE$+s(e)$ procedure requires the knowledge of $e_0(h)$. 
We define $e_{00} = e_0(h=0)$.
As no numerical method is currently able to give it to the required precision, we browse a range of values and select the one that gives the most coinciding results for $h=0$\cite{suppMat}. 
This leads to values near the ones infered from DMRG ($e_{00}=-0.4386(5)$)\cite{Yan2011, DMRG_kagome, PhysRevB.100.155142} and exact diagonalization ($e_{00}=-0.4387039$ for a 48 site cluster)\cite{PhysRevB.100.155142, Wietek2018}.

For small $h\neq0$, the energy is given by 
\begin{align}
e_0(h) &\simeq e_{00} - \frac 12 \chi_0 h^2,
\end{align}
as thermodynamic relations imply that $\chi_0=\chi(T=0,h=0)=-\frac{d^2 e_0(h)}{dh^2}$. 
While $\chi_0$ is 0 in gapped systems, as the ground state remains unchanged for infinitesimal $h$, we a priori have $\chi_0\neq0$ for gapless systems. 
To give an idea of the $\chi_0$ value, we can look at the classical model\cite{PhysRevLett.88.057204}, which is gapless: $\chi_0=S/6$. 
A recent ED study\cite{SAKAI201885} uses the energy in different spin sectors and for different lattice sizes to get a possibly finite $\chi_0$, which is also compatible with sine-square deformation results\cite{PhysRevB.98.140405}. 
We choose here to consider $\chi_0$ as an input parameter and to deduce $e_0(h)$ from $e_{00}$ and $\chi_0$. 
Another possibility is to self-determine $e_0(h)$ and to extract $\chi_0$ from it, but this is not conclusive. 
Indeed, our procedure (see Sec.~II.E of\cite{suppMat}) allows a determination of $e_0(h)$ with some uncertainties and $\chi_0$, being related to the second derivative of $e_0(h)$, suffers for even much larger uncertainties; therefore, we find that almost any reasonable value of $\chi_0$ is compatible with our results.

We note $s'$ and $s''$ the derivatives of $s$ with respect to $e$ at constant $h$. 
We recall that $\beta=s'$. 
The specific heat per site $c_V$ and magnetization per site $m$ are:
\begin{align}
\label{eq:cV_m}
c_V=-\frac{s'^2}{s''},\qquad
m= \frac{1}{s'}\left.\frac{\partial s}{\partial h}\right|_e.
\end{align}
We emphasize that $m$ is now obtained directly from $s(e,h)$, simplifying the procedure used in\cite{Bernu2015}. 
We deduce from $m$ the experimentally measured magnetic susceptibility per site $\chi=m/h$. 

\begin{figure}[t]
    \includegraphics[width=.45\textwidth]{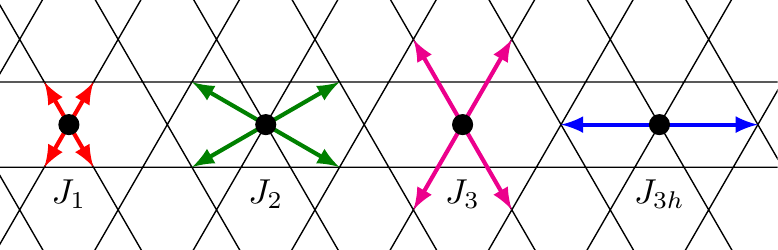}
    \caption{
        First ($J_1$), second ($J_2$) and third ($J_3$ and $J_{3h}$) neighbor interactions on the kagome lattice.  
        \label{fig:neigh_kag}
    }
\end{figure} 

At the end of the day, for a given spin model, we extrapolate $\chi(T)$ and $c_V(T)$ at all temperatures from the HTSE, with, as supplementary input, the values of $e_0$, $\chi_0$, and $\alpha$. 
Fig. \ref{fig:Cv_chi} shows $c_V$ and $\chi$ for the unperturbed Hamiltonian of Eq.~\eqref{EQ-H0}.
The assumption on $\alpha$ has no influence for $T>0.3$: HTSE strongly constrain the functions in this domain of temperature. 
Notably, the high temperature peak of $c_V$ near $T=0.7$ is well determined, which is not the case for the small temperature secondary peak ($T\simeq0.03$). 
The existence of such a peak or shoulder, sign of a large amount of low energy states, is still highly debated as it is very sensible to eventual finite size effects\cite{PhysRevLett.111.010401, PhysRevB.98.094423}. 

At this point, it is important to emphasize a particularity of the KHAF. 
In most of simpler models, we are not able to get convincing results if we arbitrarily chose $\alpha$ or $e_0$, in the sense where we do not get several mingled PAs for $G_\alpha(e,h)$: only physically correct hypothesis give a collection of coincinding PAs. 
In this respect, KHAF is very special as any hypothesis on $\alpha$ leads to valuable extrapolations: no $\alpha$ can be discarded by this way.

\section{Results for the modified KHAF}
\label{sec:perturbations}

We now add different terms to $\H_0$, whose effects will be studied successively below:
\begin{eqnarray}
\label{eq:Ham_pert}
\H &=& \H_0
- h \sum_iS^z_i
+ \sum_{\langle i,j\rangle} \left(D_z\cdot(\mathbf S_i\land \mathbf S_j)_z +
\delta_z S^z_iS_j^z\right)\\
&&+ J_2 \sum_{\langle i,j\rangle_2} \mathbf S_i\cdot \mathbf S_j
+ J_{3} \sum_{\langle i,j\rangle_3} \mathbf S_i\cdot \mathbf S_j
+ J_{3h} \sum_{\langle  i,j\rangle_{3h}} \mathbf S_i\cdot \mathbf S_j,
\nonumber
\end{eqnarray}
where $D_z$ is the $z$ component of the DM vector, $\delta_z$ the Ising anisotropy, $J_2$, $J_3$ and $J_{3h}$ the second and third nearest-neighbor terms (see Fig.~\ref{fig:neigh_kag}).
The $Q_{n,k}$ of Eq.~\eqref{eq:lnZ} are now polynomials of order $n$ in the rate of vacancies $p$, $D_z$, $\delta_z$, $J_2$ and $J_{3h}$. 
The HTSE order depends on the complexity of the Hamiltonian: order 20 is obtained for the KHAF with impurities, 18 with the Ising anisotropy, 16 with DM interactions, 15 with second and third neighbor exchanges.

\begin{figure}[t]
	\includegraphics[width=0.48\textwidth]{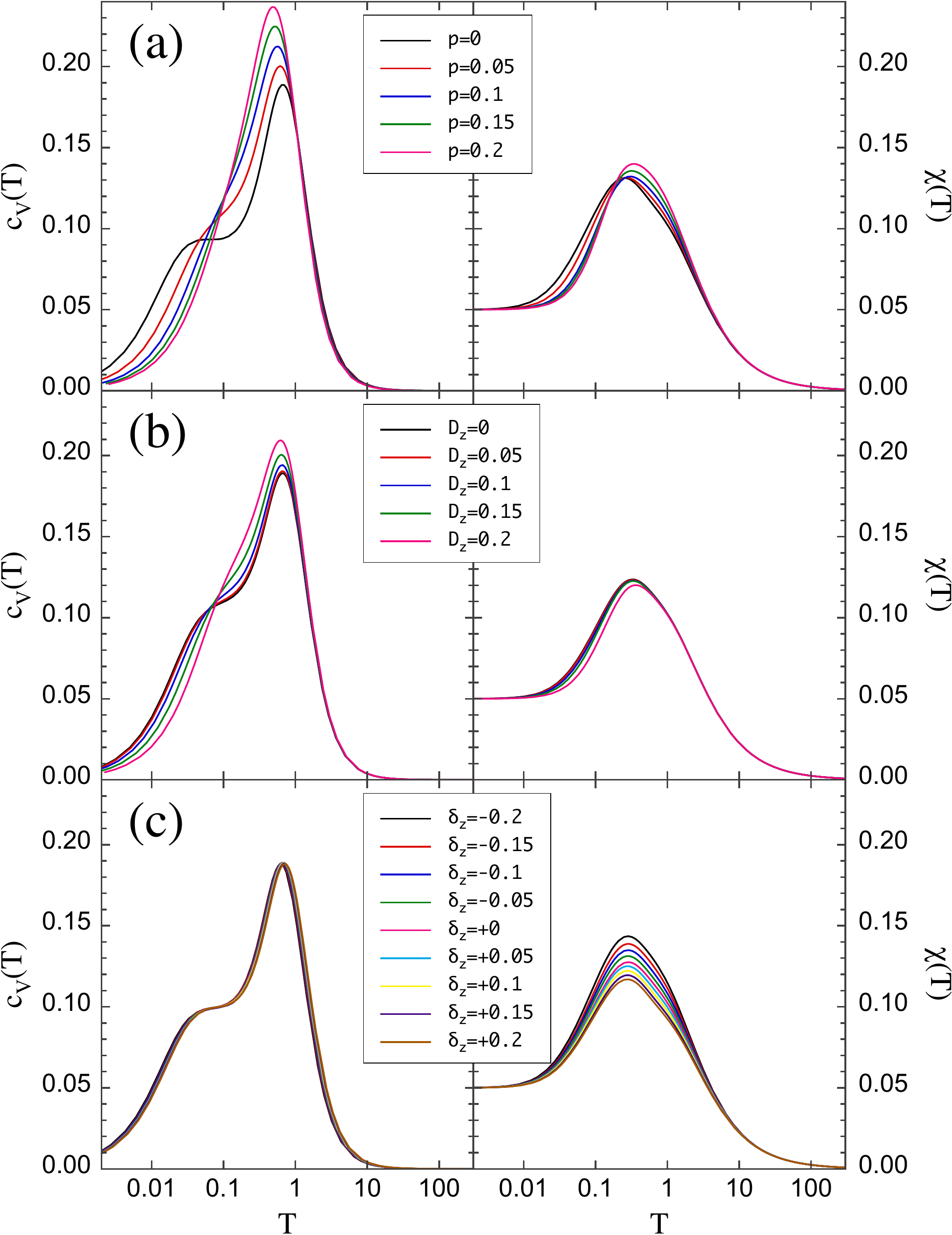}
	\caption{HTSE$+s(e)$ results on the KHAF: specific heat $c_V$ and magnetic susceptibility $\chi$ for different (a) vacancy rates $p$, (b) DM interactions $D_z$ and (c) Ising anisotropy $\delta_z$. 
		Results for $\alpha=1$ (linear low $T$ $c_V$) and for $\chi_0=0.05$ are shown. 
		The results for $\alpha=0$ or 2 and for other $\chi_0$ are in\cite{suppMat}.}
	\label{fig:perturbations}
\end{figure}

\begin{figure}[t]
	\includegraphics[width=.33\textwidth]{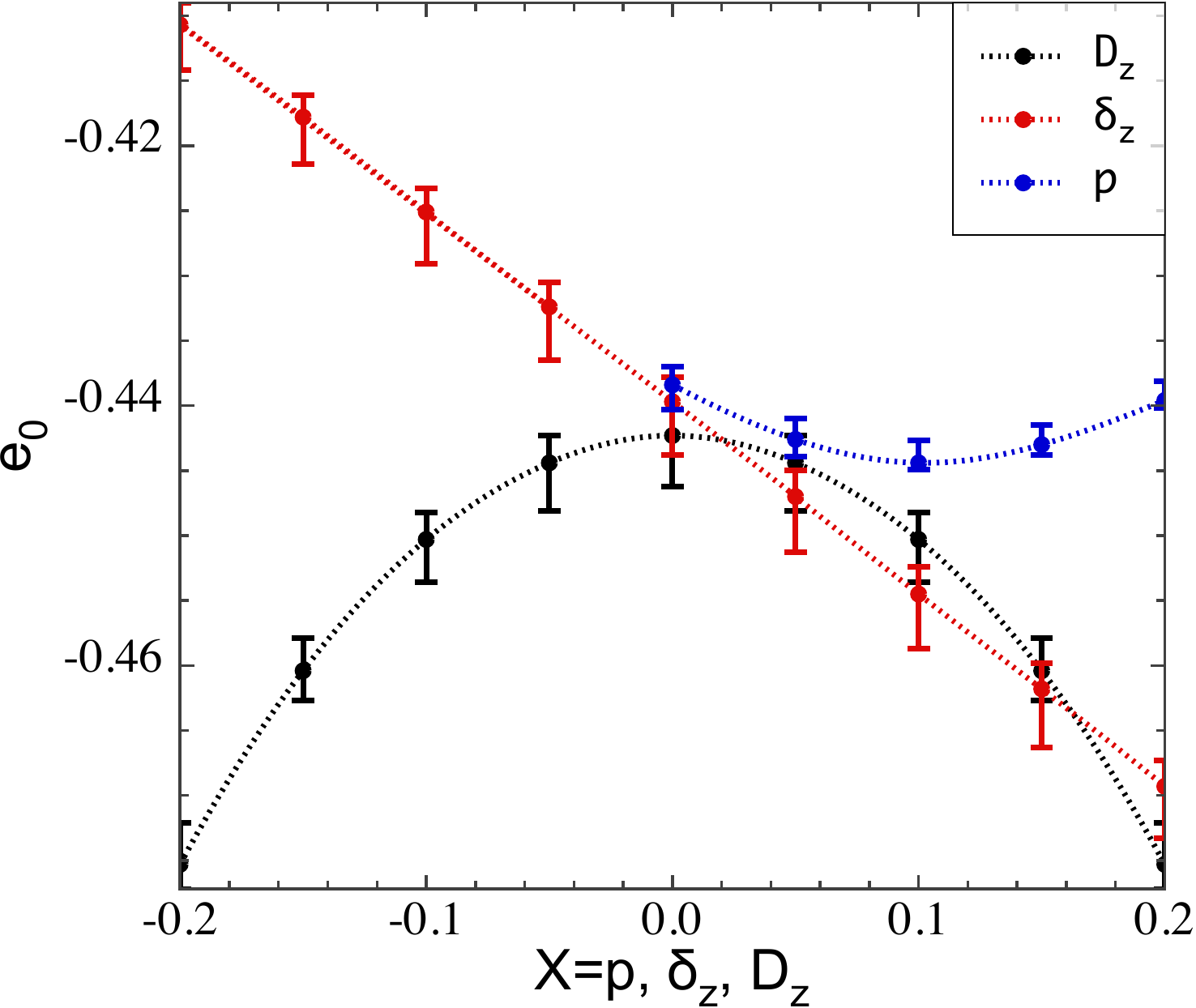}
	\caption{Ground state energies $e_0$ for different impurity rates $p$, DM interaction strength $D_z$ and Ising anisotropy $\delta_z$, for $\alpha=1$.
		The results differ at $X=0$ (pure KHAF) due to the different HTSE orders used for the various types of perturbations $X$. }
	\label{fig:energy}
\end{figure}

Fig.~\ref{fig:perturbations} shows the influence on $c_V$ and $\chi$ of some of these perturbations and of a dilution rate, for $h=0$ and with the hypothesis that $\alpha=1$. 
To get $\chi$, we need a supplementary hypothesis on $\chi_0$, chosen to be 0.05 for this figure (results for other values are in \cite{suppMat}). 
Note that at intermediate temperatures, our results are consistent with Numerical Linked Cluster (NLC) results\cite{DM_cluster, PhysRevB.76.184403}. 
Fig.\ref{fig:energy} shows how the ground state energy, $e_0$, extracted from the most coinciding HTSE$+s(e)$ extrapolations\cite{suppMat}, evolves with the considered perturbations. 

{\bf Impurities.} 
The rate of vacancies (magnetic Cu replaced by non magnetic Zn atoms) in the kagome lattice of Herbertsmithite is experimentally estimated to be less than $5\%$\cite{Mendels}. 
We suppose here that interactions between remaining spins are unchanged. 
The extracted $e_0(p)$ has a minimum around $p=10\%$ (Fig.~\ref{fig:energy}). 
For classical spins, a low $p$ does not modify the energy per \textit{magnetic site}\cite{PhysRevLett.70.3812} (even if it lowers the energy per \textit{lattice site}) and this minima cannot be reproduced. 
But for quantum $1/2-$spins \cite{PhysRevLett.104.177203, DM_impurity, PhysRevB.68.224416}, it can be qualitatively understood as the minimal energies $E_t$ on a triangle and $E_b$ on a bond are the same ($-3/4$), whereas classically, $E_t<E_b$ ($-3S^2/2$ against $-S^2$). 
A rough approximation of the energy per spin on the lattice is 
$$\frac{2(1-p)^2}{3}E_t+2(1-p)pE_b$$
 and reproduces the minimum at $p\simeq10\%$ if $E_t\simeq 4 E_b/3$, which seems reasonable. 

At finite temperature, we find that impurities soften the separation of the two peaks in $c_V$, strengthen $\chi$ and shift it to higher temperatures (Fig.~\ref{fig:perturbations}(a)). 
Another type of defects is present in Herbertsmithite but not treated here: interlayer magnetic atoms (Zn replaced by Cu atoms) at a rate of $15\%$ of occupation\cite{Mendels, refId0}. 
They will enforce the tridimensional character of the compound. 

{\bf \DM interaction.} 
This interaction originates from the spin-orbit coupling\cite{Dzyaloshinskii, Moriya} and is often considered, in Herbertsmithite, as the main deviation from the KHAF, together with impurities\cite{DM_impurity}. 
The out of plane component $D_z$ is supposed to be dominant and is the only one considered here.
The combined effect of the in and out of plane $\mathbf D$ has been studied by NLC\cite{DM_cluster, PhysRevB.76.184403}. 
The sum in the Hamiltonian \eqref{eq:Ham_pert} is over oriented links, all pointing in the same arbitrary direction when we turn around the lattice hexagons.
In Herbertsmithite, $D_z\simeq 0.04\,J_1$\cite{DM_Bert}. 
Order is supposed to appear for $D_z\simeq0.08\,J_1$\cite{DM_Cepas, Messio_DM, PhysRevLett.118.267201}, even if smaller values ($D_z\simeq0.01\,J_1$) have recently been proposed\cite{PhysRevB.98.224414}. 
We find that $D_z$ enhances the main $c_V$ peak and has a weak effect on $\chi$ (Fig.~\ref{fig:perturbations}(b)). 
As expected, $e_0(D_z)$ behaves quadratically (Fig \ref{fig:energy}). 

{\bf Ising anisotropy.} 
The Ising anisotropy $\delta_z$ interpolates between the ferromagnetic Ising model ($\delta_z=-\infty$), the $XY$ model ($\delta_z=-1$) and the antiferromagnetic Ising model ($\delta_z=\infty$), staying in the same spin liquid phase for $\delta_z>-1$\cite{PhysRevLett.114.037201}. 
Moreover, an exactly solvable point $\delta_z=-3/2$ was recently discovered and analyzed\cite{PhysRevB.99.104433}.
For small $\delta_z$, $e_0(\delta_z)$ is linear (Fig \ref{fig:energy}). 
This can be qualitatively understood by considering that most of the energy contribution in the ground state comes from the concentration $c$ of singlet bonds, whose energy is $-(3+\delta_z)/4$. 
With this naive picture, we get $e_0(\delta_z)=e_0(\delta_z=0)(1+\delta_z/3)$, whose slope is in agreement with the one fitted from HTSE data $0.146(1)\sim -0.44/3$ (Fig.~\ref{fig:energy}). 
Similarly, the susceptibility of such singlets decreases when $\delta_z$ increases and reciprocally, which is the behavior seen in Fig.~\ref{fig:perturbations}(c). 
On the contrary, $c_V$ is almost insensitive to $\delta_z$.  

{\bf Second and third neighbors interactions.} 
$J_2$, $J_3$ and $J_{3h}$ are known to lift the classical degeneracy of the KHAF toward the $\sqrt3\times \sqrt3$ long range order for $J_2<0$ and $J_3>0$, towards the $q=0$ order for $J_2>0$, $J_3<0$ and $J_{3h}<0$ and towards the cuboc1 order for $J_{3h}>0$\cite{Regular_order, PhysRevLett.118.267201}. 
For quantum spins 1/2, small changes in these parameters have seemingly low influence and preserve the spin liquid phase for $|J_2|, |J_{3h}|\leq0.2$\cite{PSG_chiral_fermions, PhysRevB.91.104418, PhysRevB.91.075112, PhysRevB.91.020402}. 
The $J_3$ case is less studied. 
These terms add new links to the KHAF model, therefore HTSE are limited to order 15. 
$J_2$ and $J_3$ have stronger effects  than $J_{3h}$ on $c_V(T)$ and $\chi(T)$.
Results are displayed and discussed in \cite{suppMat} for completeness. 

\begin{figure}[t]
	\includegraphics[width=.35\textwidth]{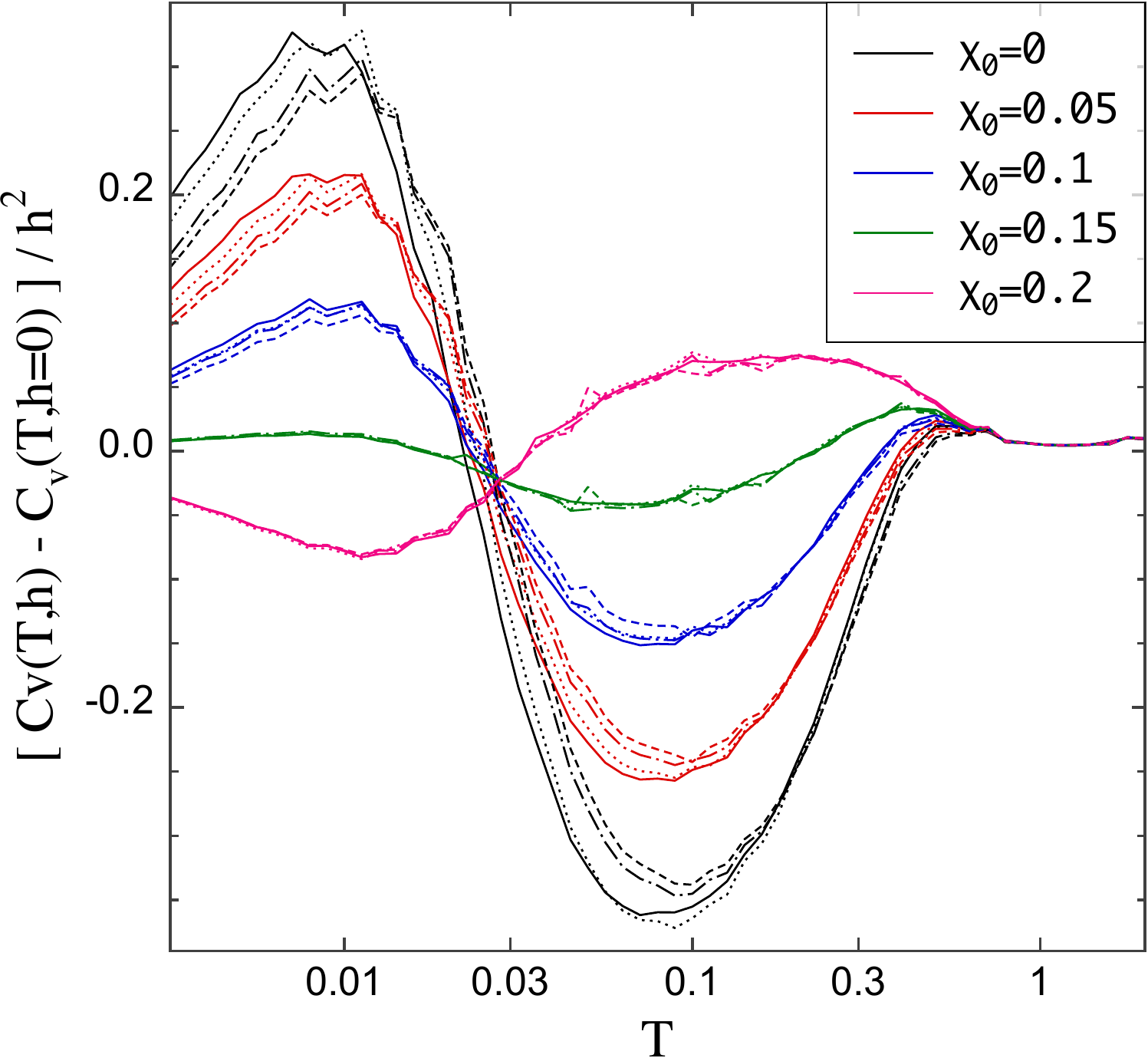}
	\caption{Variation of the specific heat $c_V(T)$ using various values of $\chi_0$. 
    Shown are $(c_V(T,h)-c_V(T,h=0) )/h^2$ for $\alpha=1$ and several values of $h$: 0.2 (full lines), 0.15 (dotted lines) 0.1 (dahed-dotted lines) and 0.05 (dashed lines).
	By construction, the integral of each curve is 0. 
	For  $\chi_0=0.15$, the curve is almost flat.
	For $\chi_0<0.15$, there is an increase of  $c_V(T)$ at $T\lesssim0.03$ and a decrease for $0.03\gtrsim T\gtrsim0.3$, and the opposite for $\chi_0>0.15$.
    }
	\label{fig:h}
\end{figure}

{\bf Magnetic field.}
We now consider the effect of a magnetic field $h$, that is a special perturbation as it is easily tunable experimentally, contrarily to the previous ones. 
Up to now, HTSE coefficients were only computed at the lowest order in $h$ but are here exact. 
In a gapless system, the ground state magnetization continuously increases up to a critical field $h_c$, above which the phase changes, either towards the fully magnetized state, or towards an intermediate phase. 
For classical spins\cite{PhysRevLett.88.057204}, $h_c=2S$ at $T=0$, giving rise to the finite $T$ $\frac13$-magnetization plateau, but quantum studies\cite{ncomms3287} find a lowest $\frac19$-magnetization plateau for $h_c\simeq0.6S$. 
Thus, we focus on $h \lesssim 0.2$. However, for Herbersmithite where $J\simeq 180K$, $h\sim0.1$ is a hardly achieved field for experimentalists.
On Fig.~\ref{fig:h}, the difference $(c_V(T,h)-c_V(T,h=0))/h^2$ appears to be weakly dependent on $h$, but roughly proportional to the difference $\chi_0-0.15$. This is an interesting effect, that could be used to get an hint on the $\chi_0$ value as the phonons contributions, known to spoil the $c_V$ measurements, are a priori suppressed in this difference. 

\section{Conclusion}
\label{sec:conclusion}
In this paper, the HTSE coefficients of antiferromagnetic 1/2-spins on the kagome lattice have been exactly obtained as polynomials of various Hamiltonian parameters, with at least three more terms than previously. 
The entropy method (HTSE$+s(e)$) has been applied to these models.
Two types of gapless spin liquids (linear and quadratic low $T$ specific heat) have been considered and several values of $\chi_0$ have been explored.
We have studied the effect on $c_V$ and $\chi$ of various perturbations of the KHAF: magnetic field, impurities, DM interaction, Ising anisotropy, further neighbor couplings.
The ground state energies have been extracted with a procedure based on the number of coinciding PAs detailed the supplemental material\cite{suppMat}, leading to coherent results down to small temperatures. 

The variations of $c_V(T)$ and $\chi(T)$ are sensible to Hamiltonian perturbations below $T\sim J_1/10$. 
For Herbertsmithite, this is precisely in this range of temperature that the experimentalists get more and more precise data, therefore HTSE$+s(e)$ is a powerful tool to determine the values of the Hamiltonian parameters from them, as already demonstrated for other models. 
We notably enlightened a way to probe $\chi_0$ using $c_V$ measurements at finite $T$ under a magnetic field. 
In a near future, we expect that measurements under pressure of kagome compounds will tune some other Hamiltonian parameters, and that the impurity rate will be controlled. 

We have here treated in great details the controversial case of the KHAF, but our extrapolation technic can as well treat any statistical model if the HTSE coefficients are known. 
Our code calculating the HTSE coefficients works for spin models on any lattice, for any interaction preserving the total magnetization along $z$ and for $S=1/2$. 
What has been chosen as perturbative parameters in this paper can be set to any arbitrary value as the HTSE coefficients are exact polynomials of them. 
However, the convergence properties of the series are affected by possible phase transitions.

\textbf{Acknowledgments}
This work was supported by the French Agence Nationale de la Recherche under Grants No. ANR-18-CE30-0022-04 ’LINK’, and by the Idex \textit{Sorbonne Universit\'e} through the \textit{Emergence} program. 
The authors thank Sylvain Capponi for discussions and results on exact diagonalizations, Fran\c cois Delyon for many animated discussions. 
L. M. thanks Johannes Richter and Chisa Hotta for discussions at Physikzentrum Bad Honnef. 

\bibliography{Article_Kag_HTE}

\end{document}